\documentclass[prb,letterpaper,aps,floatfix,twocolumn]{revtex4-2}

\usepackage{graphicx}
\usepackage{amsmath,amssymb}
\usepackage{bm}

\usepackage[pdfstartview=FitH,breaklinks=true,bookmarks=true,colorlinks=true,anchorcolor=black,citecolor=red,filecolor=black,menucolor=black,urlcolor=blue,linkcolor=blue]{hyperref}

\usepackage{color}

\begin{document}

\title{Work statistics and thermal phase transitions}

\author{Kwai-Kong Ng}
\email{kkng@thu.edu.tw}
\affiliation{Department of Applied Physics, Tunghai University, Taichung 40704, Taiwan}
%
%
\author{Min-Fong Yang}
\email{mfyang@thu.edu.tw}
\affiliation{Department of Applied Physics, Tunghai University, Taichung 40704, Taiwan}

\date{\today}

\begin{abstract}
The investigation of nonequilibrium thermodynamics in quantum many-body systems underscores the importance of quantum work, which differs from its classical counterpart due to its statistical nature. Recent studies have shown that quantum work can serve as an effective indicator of quantum phase transitions in systems subjected to sudden quenches. However, the potential of quantum work to identify thermal phase transitions remains largely unexplored. In this paper, we examine several types of thermal phase transitions in a sudden-quench hard-core boson model, including Ising, three-state Potts, and Berezinskii-Kosterlitz-Thouless transitions. Through finite-size scaling analysis, we conclude that work statistics can also characterize the critical behaviors of thermal phase transitions in generic many-body systems. Our investigation paves the way for applying work statistics to characterize critical behavior in many-body systems, with implications that may extend to broader contexts.
%
\end{abstract}

\maketitle

\section{INTRODUCTION}

Investigating nonequilibrium thermodynamic properties of quantum many-body systems is a complex and demanding task. Quantum work, a concept rooted in the principles of quantum mechanics and thermodynamics, has emerged as a crucial area of study for understanding non-equilibrium quantum systems. Unlike its classical counterpart, quantum work is a fluctuating quantity that requires a statistical approach. This has led to the development of the two-point measurement scheme, where work is defined as the difference in energy between initial and final measurements on a quantum system~\cite{Talkner_etal2007}. All the available statistical information about quantum work $W$ is contained in the distribution function $p(W)$ defined for work statistics.

Not only being important for understanding fundamental thermodynamics, quantum work also provides valuable insights into quantum phase transitions (QPTs), where a system's ground state changes qualitatively due to variations in controlling parameters~\cite{Silva2008,Mascarenhas_etal2014,Sharma-Dutta2015,%
Campbell2016,Paganelli-Apollaro2016,Fei_etal2020,Zawadzki_etal2020,%
Varizi_etal2020,Zhang-Quan2022,Gu_etal2022,Zawadzki_etal2023,Lin-Huang2024}.
(For a recent review, see Ref.~\cite{review2018}.)
In this context, researchers have particularly focused on the sudden-quench scenario, wherein the system parameters are subjected to an abrupt change. It is established in Ref.~\cite{Mascarenhas_etal2014} that, analogous to the latent heat of classical phase transitions, the average work $\langle W\rangle=\int d W\,p(W)\,W$ done per quench exhibits a discontinuity at first-order QPT points. For second-order QPTs, the average work $\langle W\rangle$ changes continuous at the transition points. In contrast, particularly for weak sudden quenches, the irreversible work $\langle W_\mathrm{irr}\rangle=\langle W\rangle-\Delta F$ (with $\Delta F$ being the free energy difference after and before quench) diverges at the critical point~\cite{Mascarenhas_etal2014}. Thus, both average work and irreversible work can serve as indicators for revealing QPTs without requiring prior knowledge of order parameters or symmetries. Furthermore, analyzing the scaling behavior of irreversible work can aid in understanding the universality classes of QPTs~\cite{Sharma-Dutta2015,Paganelli-Apollaro2016}. These conclusions arise from the close relationship between average work and the first derivative of ground-state energy with respect to the quench parameter, and between irreversible work and the second derivative. The validity of quantum work in characterizing the quantum criticality of many-body systems has been corroborated by previous studies across various systems~\cite{Silva2008,Mascarenhas_etal2014,Sharma-Dutta2015,%
Campbell2016,Paganelli-Apollaro2016,Fei_etal2020,Zawadzki_etal2020,%
Varizi_etal2020,Zhang-Quan2022,Gu_etal2022,Zawadzki_etal2023,Lin-Huang2024,review2018}. 

Most early investigations focused exclusively on QPTs, which take place only at or near absolute zero temperature. This is because thermal fluctuations generally obscure the nonanalytic behaviors associated with QPTs, thereby diminishing the effectiveness of work statistics in characterizing the criticality of many-body systems at finite temperatures. Nevertheless, the authors in Ref.~\cite{Zhang-Wu2022} challenge this notion. They examine two solvable models, the Dicke model and the Lipkin-Meshkov-Glick model, from which analytic expressions for the average work $\langle W\rangle$ in the thermodynamic limit can be derived. They find that $\langle W\rangle$ displays a cusp at thermal phase transitions, indicating a nonanalyticity in its derivatives that arises solely from thermal fluctuations. Their conclusions suggest that quantum work statistics can also be employed to identify thermal phase transitions.

However, generic many-body systems are usually not analytically solvable, making numerical calculations on finite-sized systems the only feasible approach. Due to finite-size effects, the cusps in the average work of limited-sized systems may become smeared. Consequently, it raises the question of whether a precise determination of thermal phase transition points can still be attained using this approach.

In this paper, we address this issue by examining a hard-core boson model on both square and triangular lattices with nearest-neighbor hopping and repulsion~\cite{Batrouni-Scalettar2000,Hebert_etal2001,Schmid_etal2002,%
Wessel-Troyer2005,Heidarian-Damle2005,Melko_etal2005,Boninsegni-Prokofev2005}. The models under consideration can exhibit several types of thermal phase transitions, including melting transitions characterized by Ising or three-state Potts universality, as well as normal-superfluid transitions of the Berezinskii-Kosterlitz-Thouless (BKT) type~\cite{Schmid_etal2002,Boninsegni-Prokofev2005}. Here, we employ the widely used quantum Monte Carlo (QMC) method known as the stochastic series expansion (SSE) algorithm~\cite{QMCreview} to measure the average work $\langle W\rangle$ and its temperature derivative $\partial\langle W\rangle/\partial T$. As discussed in the Appendix, the SSE method provides a direct estimate of $\partial\langle W\rangle/\partial T$ based on the operator sequence from the simulation, thereby avoiding any additional noise that could arise from numerically differentiating the average work. Through the analysis of finite-size scaling~\cite{Goldenfeld,Cardy}, we demonstrate that the temperature derivative of quantum work can effectively detect and differentiate between different types of second-order thermal phase transitions and accurately locate their critical points. More interestingly, we find that this approach can also be applied to infinite-order BKT transitions~\cite{Berezinskii1971,Kosterlitz-Thouless1973}. Our results thus confirm the general validity of quantum work in characterizing the criticality of many-body systems.

The remainder of this paper is organized as follows.
The relationship between average work and free energy for systems subjected to sudden quenches is discussed in Sec.~\ref{sec:Q_work}. We then derive the finite-size scaling for the average work and its derivative with respect to temperature.
In Sec.~\ref{sec:results}, we introduce our model and the characteristics of the thermal phase transitions it encompasses. We then present our QMC results for the average work and its temperature derivative associated with these transitions.
We conclude our paper in Sec.~\ref{sec:conclusion}.
The Appendix explains how the temperature derivative of average work can be measured directly based on the operator sequence from the simulation.

\section{quantum work around thermal phase transitions}\label{sec:Q_work}

\subsection{Average work and its temperature derivative}\label{sec:avg_work}

Unlike some conservative quantities, quantum work is not an observable and will depend on the microscopic details/paths of the non-equilibrium process. Due to its stochastic nature, the work performed on a finite system is described by a distribution function defined by the following two-point measurement scheme~\cite{Talkner_etal2007}.

Let us consider a quantum system described by the Hamiltonian $H(\lambda)$ with $\lambda$ being a time-dependent externally controllable parameter. At the beginning, the system with an initial value $\lambda$ is prepared in a mixed state $\rho_0=e^{-\beta H(\lambda)}/\mathrm{Tr}[e^{-\beta H(\lambda)}]$ in equilibrium with a heat bath at an inverse temperature $\beta=1/T$. The initial thermal equilibrium state is then decoupled from the bath and the controlling parameter is tuned to its final value $\lambda^\prime=\lambda+\delta$, such that the subsequent time evolution of the system is entirely dictated by a unitary operator $\hat{U}$. The distribution function $p(W)$ of work $W$ at the final time, which encodes the full statistics of work, is expressed as~\cite{Talkner_etal2007}
\begin{equation}
p(W)=\sum_{m,n} \delta\left(W-[E_m(\lambda^\prime)-E_n(\lambda)]\right) \, p(m|n) \;p_n \;.
\end{equation}
Here $p(m|n)=|\langle m(\lambda^\prime)|\hat{U}|n(\lambda)\rangle|^2$ is the transition probability from the energy eigenstate $|n(\lambda)\rangle$ of energy $E_n(\lambda)$ at the initial time to another eigenstate $|m(\lambda^\prime)\rangle$ of energy $E_m(\lambda^\prime)$ at the final time. $p_n=\langle n(\lambda)|\rho_0|n(\lambda)\rangle$ is the distribution function of the initial thermal equilibrium state.
The differences $E_m(\lambda^\prime)-E_n(\lambda)$ between the outcome of an energy measurement performed in the eigenstates of the final Hamiltonian and the initial energy of the system gives the work done on the system during the above nonequilibrium process.

For an instantaneous quench from $\lambda$ to $\lambda^\prime=\lambda+\delta$, we have $\hat{U}\to1$. In this case of sudden quench, the average work performed becomes
\begin{align}
\langle W\rangle &=\int d W\,p(W)\,W \nonumber \\
&=\sum_{m,n} [E_m(\lambda^\prime)-E_n(\lambda)] \,
|\langle m(\lambda^\prime)|n(\lambda)\rangle|^2\,
p_n \nonumber \\
&=\mathrm{Tr}\left[H(\lambda+\delta)\rho_0\right] - \mathrm{Tr}\left[H(\lambda)\rho_0\right] \; . \label{avg_work2}
\end{align}
In the present study, we focus on the quantum systems with the Hamiltonian  $H(\lambda)=H_0+\lambda\,H_1$. Therefore, $H(\lambda+\delta)-H(\lambda)=\delta\,[\partial H(\lambda)/\partial\lambda]$ and the average work thus becomes directly related to the first derivative of free energy:
\begin{equation}\label{avg_work}
\langle W\rangle
=\delta\,\frac{\partial F(\lambda)}{\partial\lambda} \; ,
\end{equation}
where $F(\lambda)=-(1/\beta)\ln\mathrm{Tr}[e^{-\beta H(\lambda)}]$ is the free energy of the initial thermal equilibrium state. We note that the above discussions hold generally, regardless of the value of $\delta$. Thus, similar to studies of quantum phase transitions at zero temperature~\cite{Mascarenhas_etal2014}, the average work exhibits a discontinuity at first-order thermal phase transition points, where the free energy shows a cusp. This indicates that the average work can effectively detect first-order thermal phase transitions at which nonanalyticity in the first derivatives of the free energy occurs.

For second-order thermal phase transitions under consideration, since nonanalyticity in the free energy occurs only in its second (and higher) derivatives, the average work will behave continuously while exhibiting a cusp at the transition point. This fact is first revealed in Ref.~\cite{Zhang-Wu2022} using two solvable models, the Dicke model and the Lipkin-Meshkov-Glick model, in the thermodynamic limit. However, as discussed in the Introduction, the cusps in the average work of finite-sized systems may be smeared by finite-size effects, complicating the precise determination of the transition points.

Instead, we concentrate on the temperature derivative of quantum work $\partial\langle W\rangle/\partial T$, which is proportional to the second derivative of free energy and is expected to exhibit singular behavior at the second-order thermal phase transition points. We note that this quantity can be measured directly based on the operator sequence from the simulation, thereby eliminating any additional noise that may arise from numerical differentiation. However, due to finite-size effects, the divergence in $\partial\langle W\rangle/\partial T$ will manifest as local maxima or minima in practical numerical calculations. By employing the finite-size scaling for these peak/dip locations and magnitudes discussed in the next subsection, we show that the temperature derivative of the average work can identify various second-order thermal phase transitions with success. Furthermore, as discussed in Sec.~\ref{sec:BKT}, this approach can also be applied to locate the infinite-order BKT transitions with caution.

\subsection{Finite-size scaling for average work and its temperature derivative}\label{sec:fss}

Consider a system in $d$ spatial dimensions with linear dimension $L$ and volume $V=L^d$. In infinite volume and near the critical temperature $T_c(\lambda)$ of a continuous thermal phase transition for a given $\lambda$, the system is characterized by a power-law diverging correlation length $\xi\sim\left|T-T_c(\lambda)\right|^{-\nu}$, where $\nu$ denotes the correlation length exponent. According to the scaling hypothesis, the singular part of the free energy density $f_s\equiv F_s/L^d$ scales like~\cite{Goldenfeld,Cardy}
\begin{equation}
f_s \sim \xi^{-d} \sim\left|T-T_c(\lambda)\right|^{d\nu} \; .
\end{equation}
Since the shifted value $\delta$ of the quenched parameter is irrelevant in the present discussions, we will henceforth focus on the average work per lattice site per quench, defined as $\langle\mathcal{W}\rangle=(1/L^d)\langle W\rangle/\delta$, which we will refer to as the reduced average work.
From Eq.~\eqref{avg_work}, the reduced average work $\langle\mathcal{W}\rangle$ scales as
\begin{equation}
\langle\mathcal{W}\rangle \sim \frac{\partial f_s}{\partial\lambda}
\propto\left|T-T_c(\lambda)\right|^{d\nu-1}
\frac{\partial T_c}{\partial\lambda}
\propto \xi^{(1/\nu)-d} \; .
\end{equation}
This implies that the temperature derivative of the reduced average work scales as
\begin{equation}
\frac{\partial\langle\mathcal{W}\rangle}{\partial T}
\propto\left|T-T_c(\lambda)\right|^{d\nu-2}
\frac{\partial T_c}{\partial\lambda}
\propto \xi^{(2/\nu)-d} \; .
\end{equation}

For systems of finite sizes, the standard finite-size scaling hypothesis assumes that $\xi$ is bounded by the linear size $L$. Therefore, we have $L=c\left|T^*(L)-T_c\right|^{-\nu}$ ($c$ is a non-universal constant), which leads to a size-dependent pseudo-critical temperature:
\begin{equation}\label{T_c}
T^*(L)=T_c + aL^{-1/\nu}
\end{equation}
with $a=c^{1/\nu}$.
In addition, the reduced average work and its temperature derivative obey the following size scaling relations:
\begin{align}
&\langle\mathcal{W}\rangle \sim L^{(1/\nu)-d} \; , \label{scaling_W} \\
&\frac{\partial\langle\mathcal{W}\rangle}{\partial T}  \sim L^{(2/\nu)-d} \; . \label{scaling_dW}
\end{align}
This indicates that the temperature derivative of the average work will exhibit divergent behavior as $L\to\infty$ if $\nu<2/d$. In the marginal case when $\nu=2/d$, one expects a logarithmic correction to the scaling, such that the temperature derivative of the average work shows a logarithmic divergence.

The aforementioned conclusions apply to usual continuous thermal phase transitions. For the BKT transitions, there exists essential singularity in the correlation length. We thus have an exponentially rather than the usual algebraically increasing correlation length $\xi \sim \xi_0 \exp\left({c/\sqrt{T-T_\mathrm{BKT}}}\right)$ ($\xi_0$ and $c$ are non-universal constants) when approaching the critical temperature $T_\mathrm{BKT}$ from above~\cite{Kosterlitz1974,Tomita-Okabe2002,Hsieh_etal2013}. The square of $c$ can be interpreted as the width of the BKT transition. The finite-size scaling theory gives $L=\xi_0\exp\left(c/\sqrt{T^*(L)-T_\mathrm{BKT}}\right)$ such that the size-dependent pseudo-critical temperature becomes~\cite{Tomita-Okabe2002,Hsieh_etal2013}
\begin{equation}\label{T_BKT}
T^*(L)=T_\mathrm{BKT}+a/\ln^2(bL)
\end{equation}
with $a=c^2$ and $b=1/\xi_0$.

\section{numerical results}\label{sec:results}

To illustrate the ability of quantum work in detecting thermal phase transitions, we examine the two-dimensional hard-core boson model on both a square and a triangular lattice. The model Hamiltonian under periodic boundary conditions is given by
\begin{equation}\label{eq:model}
H = -t \sum_{\langle i,j \rangle} \left( b_i^{\dagger} b_j + b_j^{\dagger} b_i \right) + V \sum_{\langle i,j \rangle} n_i n_j - \mu \sum_i n_i \; ,
\end{equation}
where $b_i^{\dagger}$ ($b_i$) are the creation (annihilation) operators for hard-core bosons, $n_i=b_i^{\dagger}b_i$ is the number operator. Additionally, $t$ is the hopping parameter, $V$ represents the nearest neighbor repulsion, and $\mu$ denotes the chemical potential. The spatial dimension for our systems is $d=2$.

By adjusting the system parameters, the system at zero temperature can exist in either the solid phase or the superfluid phase~\cite{Batrouni-Scalettar2000,Hebert_etal2001,Schmid_etal2002,%
Wessel-Troyer2005,Heidarian-Damle2005,Melko_etal2005,Boninsegni-Prokofev2005}. In the case of a triangular lattice, a novel supersolid phase may also emerge~\cite{Wessel-Troyer2005,Heidarian-Damle2005,Melko_etal2005,%
Boninsegni-Prokofev2005}. The thermal melting transitions from the solid phases exhibit Ising-type universality for the square lattice case~\cite{Schmid_etal2002}, whereas the triangular lattice case displays three-state Potts universality~\cite{Boninsegni-Prokofev2005}. In contrast, the thermal transitions from the superfluid to the normal fluid follow the BKT scenario.

In the following, we calculate the reduced average work $\langle\mathcal{W}\rangle$ and its temperature derivative $\partial\langle\mathcal{W}\rangle/\partial T$ using the SSE QMC method. We emphasize that $\partial\langle\mathcal{W}\rangle/\partial T$ can be directly measured from the operator sequence obtained in the simulation (see Appendix), thereby eliminating potential numerical errors that may arise from numerical differentiation. For simplicity, we condiser the case of quenching the hopping parameter from $t$ to $t+\delta$, which results in the following expression for the reduced average work derived from Eq.~\eqref{avg_work2}:
\begin{equation}
\langle\mathcal{W}\rangle=\frac{1}{L^2}\frac{1}{t} \mathrm{Tr}(H_t\,\rho_0) \; .
\end{equation}
Here $H_t=-t\sum_{\langle i,j\rangle} b^\dagger_i b_j + \mathrm{h.c.}$ represents the kinetic term of the Hamiltonian in Eq.~\eqref{eq:model} and $\rho_0=e^{-\beta H}/\mathrm{Tr}(e^{-\beta H})$ is the density matrix of the initial thermal equilibrium state. As discussed below, by incorporating the finite-size scaling outlined in Sec.~\ref{sec:fss}, the temperature derivative of the reduced average work, $\partial\langle\mathcal{W}\rangle/\partial T$, can effectively detect all of  the aforementioned thermal phase transitions.

\subsection{Melting transition of Ising type}\label{sec:Ising}

For $V=3.0$ and $\mu=4.0$ with $t=1$ serving as the energy unit, the hard-core boson model on a square lattice, as described in Eq.~\eqref{eq:model}, resides in the checkerboard solid phase at zero temperature~\cite{Batrouni-Scalettar2000,Hebert_etal2001,Schmid_etal2002}. This phase is characterized by a broken $Z_2$ symmetry. Consequently, the symmetry-restoring melting transition at finite temperature is anticipated to be an Ising-type second-order transition, associated with a correlation length exponent of $\nu=1$~\cite{Schmid_etal2002}.

\begin{figure}[t]
\includegraphics[width=0.95\linewidth]{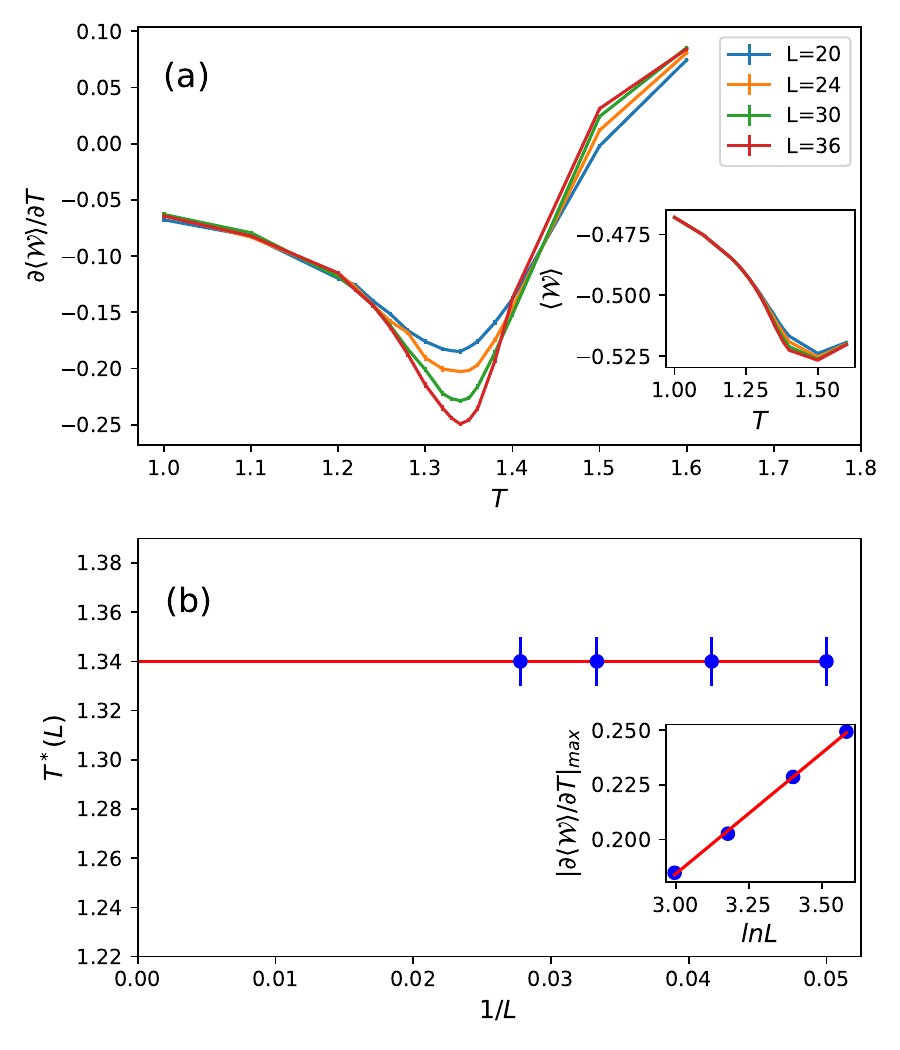}
\caption{(a) Temperature derivative $\partial\langle\mathcal{W}\rangle/\partial T$ of the reduced average work $\langle\mathcal{W}\rangle$ as a function of temperature $T$ for the hard-core boson model on a square lattice with linear sizes up to $L=36$. Here $V=3.0$ and $\mu=4.0$ with $t=1$ being the energy unit. Inset: the corresponding behavior of reduced average work $\langle\mathcal{W}\rangle$ for various sizes.
(b) The extrapolation of the size-dependent pseudo-critical temperatures $T^*(L)$ (that is, the locations of the local minima in $\partial\langle\mathcal{W}\rangle/\partial T$) by using the scaling form described in Eq.~\eqref{T_c}. $\nu=1$ is assumed for the present Ising transition. The value of the critical temperature $T_c$ in the thermodynamic limit is found to be 1.34. Inset: the size dependence of local maxima of the
absolute values $|\partial\langle\mathcal{W}\rangle/\partial T|$. The solid lines are the linear fits.
}
\label{fig:Ising_1}
\end{figure}

As illustrated in Ref.~\cite{Zhang-Wu2022}, the average work in the thermodynamic limit is expected to exhibit a cusp-like behavior, leading to nonanalyticity in its derivative. However, in the numerical calculations for systems of limited sizes, the cusps in the average work may be smeared due to finite-size effects. Fig.~\ref{fig:Ising_1} (a) displays our QMC results for the reduced average work $\langle\mathcal{W}\rangle$ and its temperature derivative $\partial\langle\mathcal{W}\rangle/\partial T$ for systems with linear sizes up to $L=36$. Both quantities vary smoothly as the temperature increases. Nevertheless, the temperature derivative shows a local minimum that becomes deeper as the system size increases. This suggests a potential divergence in $\partial\langle\mathcal{W}\rangle/\partial T$ in the thermodynamic limit. Thus, the locations of the local minima of $\partial\langle\mathcal{W}\rangle/\partial T$ should serve as size-dependent pseudo-critical temperatures  $T^*(L)$, which are anticipated to follow the scaling form given in Eq.~\eqref{T_c}. Additionally, as discussed in Sec.~\ref{sec:fss}, because $\nu=2/d=1$ for the present Ising transition, $\partial\langle\mathcal{W}\rangle/\partial T$ is expected to receive a logarithmic correction to the scaling, resulting in its minimal values displaying a logarithmic divergence. Our data presented in Fig.~\ref{fig:Ising_1} (b) support these conclusions. The extrapolated critical temperature $T_c$ is found to be 1.34, in agreement with the finding reported in Ref.~\cite{Schmid_etal2002}.

\begin{figure}[t]
\includegraphics[width=0.95\linewidth]{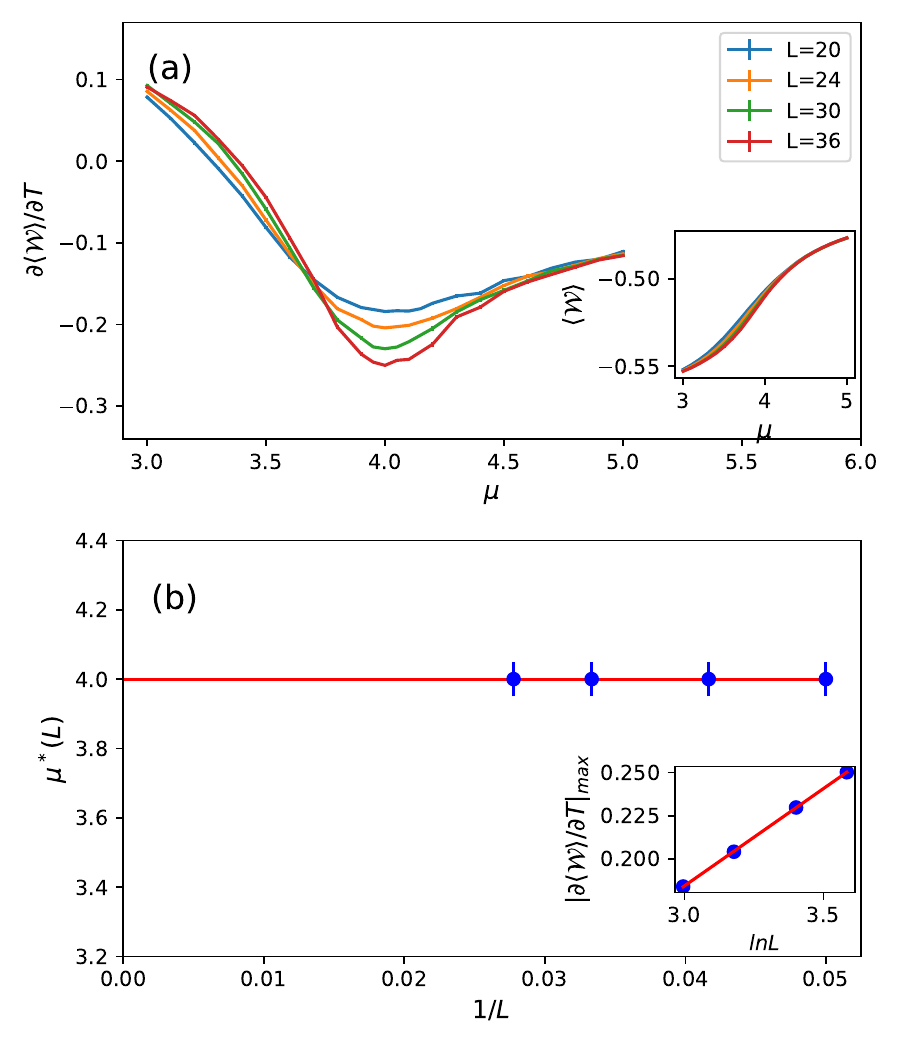}
\caption{(a) Same as Fig.~\ref{fig:Ising_1} but for varying the chemical potential $\mu$ at a fixed temperature $T=1.34$ with $V=3.0$ and $t=1$. 
(b) The extrapolation of the locations $\mu^*(L)$ of the local minima in  $\partial\langle\mathcal{W}\rangle/\partial T$, which gives the critical $\mu_c$ in the thermodynamic limit. Inset: the size dependence of local maxima of the absolute values $|\partial\langle\mathcal{W}\rangle/\partial T|$. The solid lines are the linear fits.
}
\label{fig:Ising_2}
\end{figure}

This melting transition can also be reached by varying the chemical potential $\mu$ at a fixed temperature. As a consistency check, we examine the behavior of $\partial\langle\mathcal{W}\rangle/\partial T$ at $T=1.34$, which corresponds to the critical temperature for $\mu=4$ with $V=3$ and $t=1$. As shown in Fig.~\ref{fig:Ising_2},  $\partial\langle\mathcal{W}\rangle/\partial T$ for various sizes exhibits local minima. The locations and the magnitudes of these local minima follow the expected scaling behaviors. When extrapolating the size-dependent pseudo-critical values $\mu^*(L)$ to the thermodynamic limit, they converge to $\mu_c=4.0$, which aligns with the result obtained in Fig.~\ref{fig:Ising_1}.

\subsection{Melting transition of three-state Potts universality}\label{sec:Potts}

On a triangular lattice, the zero-temperature solid phase at a rational filling $1/3$ (or $2/3$) for small $t/V$ breaks the lattice translation symmetry, resulting in a $\sqrt{3}\times\sqrt{3}$ ordering where one out of every three sites is filled (or empty)~\cite{Wessel-Troyer2005,Heidarian-Damle2005,Melko_etal2005,%
Boninsegni-Prokofev2005}. As temperature increases, the low-temperature solid phase undergoes a melting transition into a liquid phase. Due to the threefold degenerate structure of the ground state, this melting transition is anticipated to fall within the three-state Potts universality class, associated with a correlation length exponent of $\nu=5/6$~\cite{Boninsegni-Prokofev2005}.

According to the scaling relations in Eqs.~\eqref{scaling_W} and \eqref{scaling_dW}, we expect that, while the reduced average work behaves non-singularly, the temperature derivative $\partial\langle\mathcal{W}\rangle/\partial T$ diverges as $L^{2/5}$ with increasing $L$. Besides, the locations of the local extrema in $\partial\langle\mathcal{W}\rangle/\partial T$ provide the size-dependent pseudo-critical temperatures $T^*(L)$, which scale according to Eq.~\eqref{T_c} with $\nu=5/6$.

\begin{figure}[t]
\includegraphics[width=0.95\linewidth]{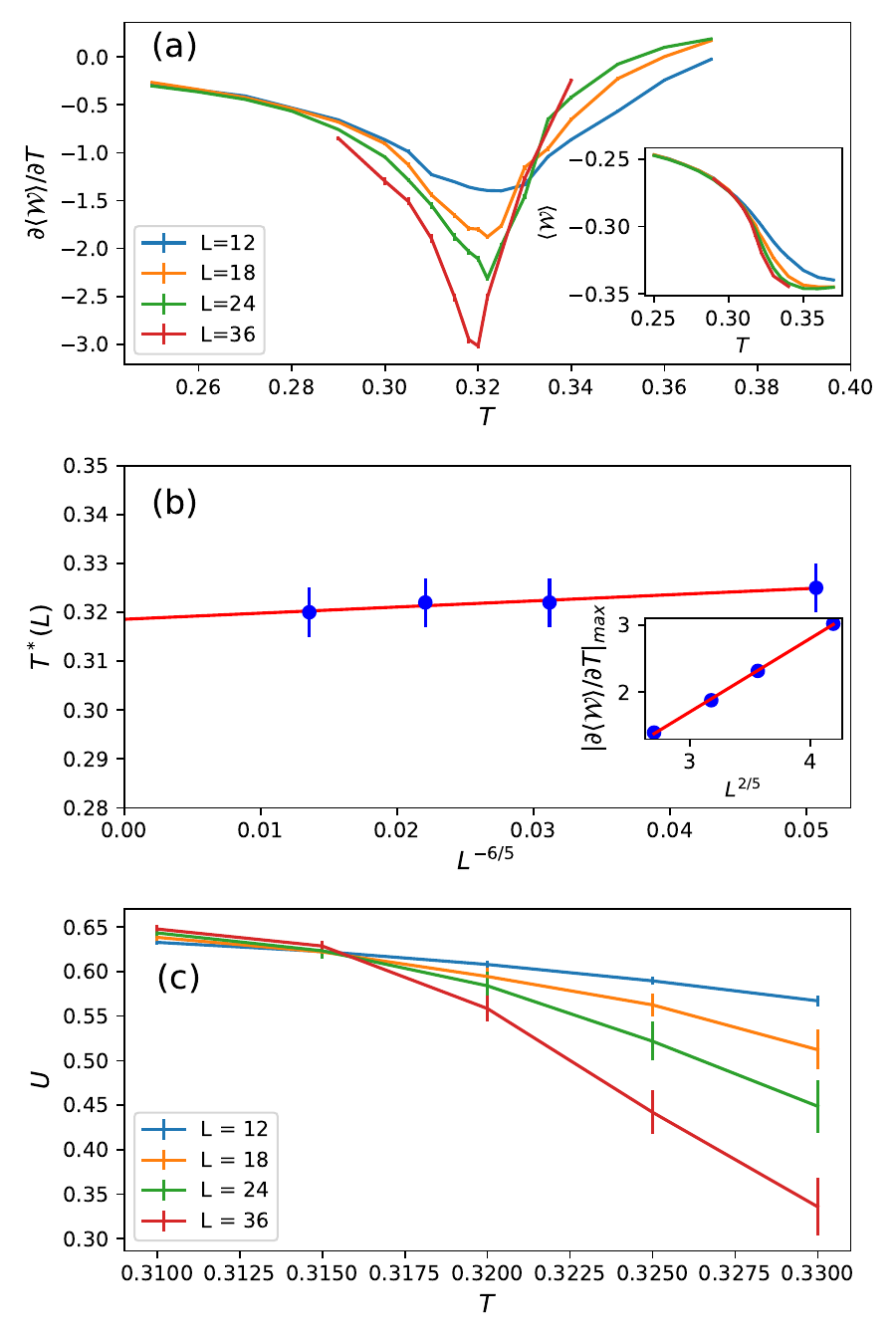}
\caption{(a) Temperature derivative $\partial\langle\mathcal{W}\rangle/\partial T$ of the reduced average work $\langle\mathcal{W}\rangle$ as a function of temperature $T$ for the hard-core boson model on a triangular lattice with linear sizes up to $L=36$. Here $t=0.1$ and $\mu=4.5$ with $V=1$ being the energy unit. The reduced average work is displayed in the inset for comparison.
(b) The size-dependent pseudo-critical temperatures $T^*(L)$ are fitted using the scaling form described in Eq.~\eqref{T_c} and the extrapolated $T_c=0.319$. Inset: the local maxima of the absolute values $|\partial\langle\mathcal{W}\rangle/\partial T|$ are fit with Eq.~\eqref{scaling_dW}. $\nu=5/6$ is assumed for this three-state Potts transition. The solid lines are the linear fits.
(c) The Binder cumulant $U$ (see the main text) for various system sizes as a function of $T$. The location of the crossing point gives the critical temperature $T_c=0.315$.
}
\label{fig:Potts}
\end{figure}

Setting $V=1$ as the energy unit, our QMC results for $t=0.1$ and $\mu=4.5$ are presented in Fig.~\ref{fig:Potts}. It can be observed that the local minima in $\partial\langle\mathcal{W}\rangle/\partial T$ indeed scale as $L^{2/5}$, indicating that they tend to diverge as the system size $L$ increases. The size-dependent pseudo-critical temperatures $T^*(L)$ are found to follow the scaling relation $T^*(L)=T_c + aL^{-6/5}$. These findings support the conclusion that the transition is characterized by three-state Potts universality with $\nu=5/6$. The critical temperature is found to be $T_c=0.319$. This value closely matches that obtained through traditional methods. For instance, as illustrated in Fig.~\ref{fig:Potts}\,(c), the crossing of the Binder cumulant $U$ gives $T_c=0.315$, lending support to the present approach. Here the Binder cumulant $U$ is defined as $U=1-\langle S(\mathbf{Q})^4 \rangle/[3 \langle S(\mathbf{Q})^2 \rangle^2]$, where
$S(\mathbf{Q})=\left| (1/N)\sum_{j=1}^{N} n_j e^{i\mathbf{Q}\cdot\mathbf{r}_j} \right|^2$ is the structure factor at the ordering vector
$\mathbf{Q}=\left( \frac{4\pi}{3}, \frac{\pi}{3} \right)$.

Our results presented in Secs.~\ref{sec:Ising} and \ref{sec:Potts} thus demonstrate the general validity of the current approach based on work statistics for detecting continuous thermal phase transitions, as long as $\nu\leq2/d$.

\subsection{Normal-superfluid transition of BKT type}\label{sec:BKT}

It is intriguing to investigate whether the infinite-order topological BKT transitions can be identified using this method. On both square and triangular lattices, the hard-core boson model at low particle (or hole) densities exhibits a superfluid phase at low temperatures. In the present two-dimensional case, the transitions between the normal and the superfluid phases at finite temperatures are expected to be of the BKT type. For illustration, we pay our attention to the square lattice case. To achieve a low-temperature superfluid phase, we set a lower chemical potential of $\mu=1.5$, keeping all other system parameters the same as those presented in Fig.~\ref{fig:Ising_1}.

\begin{figure}[t]
\includegraphics[width=0.95\linewidth]{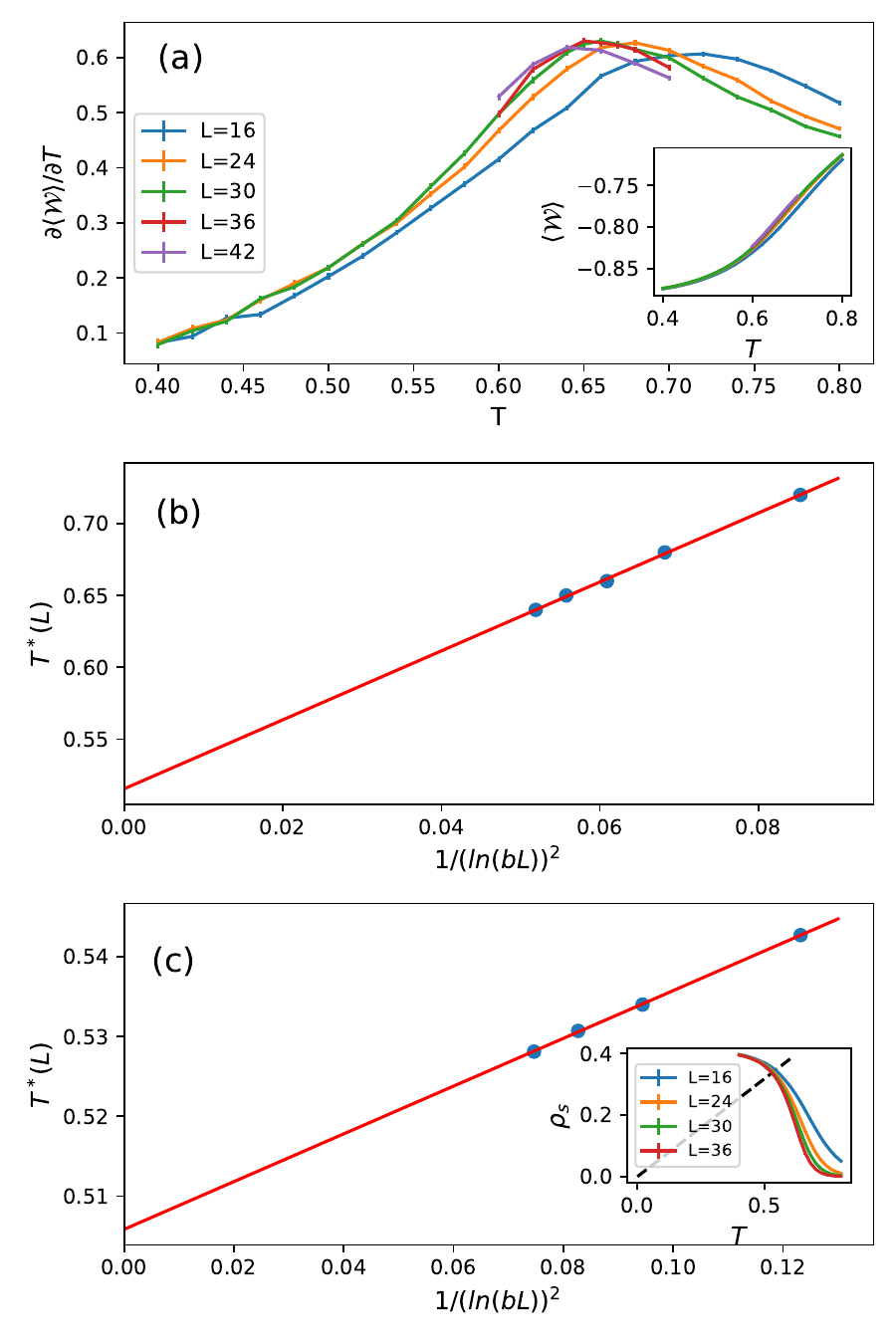}
\caption{(a) Same as Fig.~\ref{fig:Ising_1} but for lower chemical potential of $\mu=1.5$. The inset illustrates the behavior of the reduced average work.
(b) The size-dependent pseudo-critical temperatures $T^*(L)$ are fitted using Eq.~\eqref{T_BKT}. The extrapolated $T_\mathrm{BKT}=0.515$ with the fitting parameter $b=1.92$.
(c) The size-dependent pseudo-critical temperatures $T^*(L)$ determined by the Nelson-Kosterlitz universal jump condition (see the inset). Fitting with the scaling form in Eq.~\eqref{T_BKT}, we have $T_\mathrm{BKT}=0.506$ and $b=1.08$. Inset: Temperature dependence of the superfluid density for different system sizes. The dashed line denotes the critical line $2mT/\pi$ whose intersection with $\rho_s(T)$ gives $T^*(L)$ for a given $L$.
}
\label{fig:BKT}
\end{figure}

Our QMC data for the reduced average work $\langle \mathcal{W}\rangle$ and its temperature derivative $\partial\langle \mathcal{W}\rangle/\partial T$ for systems with linear sizes up to $L=42$ are presented in Fig.~\ref{fig:BKT}. In contrast to the findings in Secs.~\ref{sec:Ising} and \ref{sec:Potts}, we observe local maxima in $\partial\langle\mathcal{W}\rangle/\partial T$ here. We identify the positions of these local maxima as the size-dependent pseudo-critical temperatures $T^*(L)$. As discussed at the end of Sec.~\ref{sec:fss}, $T^*(L)$ is expected to satisfy the scaling relation given by Eq.~\eqref{T_BKT}. Our data align closely with this scaling relation, yielding $T_\mathrm{BKT}=0.515$ in the infinite-size limit.

Conventionally, the critical temperature $T_\mathrm{BKT}$ is determined using the Nelson-Kosterlitz relation~\cite{Nelson-Kosterlitz1977}, which characterizes the universal jump in the superfluid density $\rho_s$ at the BKT transition: $\rho_s=2mT_\mathrm{BKT}/\pi$. Here, $m=1/2t$ represents the effective mass of the bosons on a square lattice. In QMC simulations, the superfluid density $\rho_s(T)$ at a given temperature $T$ is calculated using the well-known winding number estimator~\cite{Pollock-Ceperley1987}. The pseudo-critical temperatures $T^*(L)$ for each system size $L$ are obtained from the intersections of the corresponding $\rho_s(T)$ values with the line defined by $2mT/\pi$. By fitting the $T^*(L)$ data to the scaling relation given in Eq.~\eqref{T_BKT}, the critical temperature $T_\mathrm{BKT}$ can be achieved. Our analysis is shown in Fig.~\ref{fig:BKT} (c), yielding $T_\mathrm{BKT}=0.506$. Remarkably, the values obtained from both the quantum work and the conventional approaches are in good agreement. This consistency supports the validity of the quantum-work approach in detecting topological BKT transitions.

Before closing this subsection, we would like to briefly comment on the subtleties related to the application of the present approach in identifying the BKT transitions. Since BKT transitions are of infinite order, the peak of $\partial\langle\mathcal{W}\rangle/\partial T$ should exhibit non-divergent behavior, as indicated in Fig.~\ref{fig:BKT}\,(a). Similar observations have been noted in the studies of specific heat~\cite{Solla-Riedel1981,Ding-Makivic1990,Ding1992,Li_etal2020} and fidelity susceptibility~\cite{Sun_etal2015,Cincio_etal2019,Zhang2021} around BKT transitions. For instance, the peak value of the specific heat is found to remain finite for a BKT transition in the thermodynamic limit, with the position of this nonsingular peak occurring at a temperature above $T_\mathrm{BKT}$.  Nevertheless, it has been shown that one can still extrapolate a value close to the BKT transition point using the scaling of the peak positions for intermediate system sizes~\cite{Sun_etal2015,Cincio_etal2019,Zhang2021}. This explains the success of our determination of $T_\mathrm{BKT}$ by employing the scaling of the peak positions of $\partial\langle\mathcal{W}\rangle/\partial T$ up to $L=42$.

\section{conclusions}\label{sec:conclusion}

In this study, we elaborate on how quantum work statistics can be utilized to characterize critical behaviors in thermal phase transitions of many-body systems undergoing a sudden quench.

For systems of finite size, we find that the temperature derivative $\partial\langle\mathcal{W}\rangle/\partial T$ of the reduced average work $\langle\mathcal{W}\rangle$ can display a local extremum around a continuous phase transition when the correlation length exponent satisfies $\nu\leq2/d$. The extremal values conform to the scaling relation presented in Eq.~\eqref{scaling_dW}, leading to singular behavior of $\partial\langle\mathcal{W}\rangle/\partial T$ in the thermodynamic limit. By employing the scaling relation in Eq.~\eqref{T_c} for the positions of these local extrema, we can accurately determine the critical temperatures $T_c$. For simplicity, we directly utilize the theoretical values of $\nu$ in our fittings. In general, $\nu$ can also be extracted through finite-size scaling, allowing us to identify not only the critical temperatures but also the universality classes of the transitions through the values of critical exponent.
Notably, our approach also enables the identification of infinite-order topological BKT transitions, where the transition temperature $T_\mathrm{BKT}$ follows the scaling relation outlined in Eq.~\eqref{T_BKT}.
As observed in Sec.~\ref{sec:results}, our findings for various types of thermal phase transitions agree well with those obtained through conventional methods, thereby affirming the validity of the present approach.

Although we focus exclusively on the hard-core boson model on square and triangular lattices, the general applicability of our approach is anticipated. A key advantage of this method is that the temperature derivative of the reduced average work serves as an indicator of thermal phase transitions without requiring prior knowledge of order parameters or symmetries within the system.
Nevertheless, this approach may not be effective for continuous transitions with $\nu>2/d$, in which case higher-order derivatives of the average work could become necessary.

In essence, our results highlight the potential of quantum work as a robust framework for investigating critical phenomena in quantum systems, thereby enhancing comprehension of both quantum and thermal phase transitions. In the present investigation, we have restricted our focus to the case of sudden quenching. Future studies could explore the effects of nonzero quench times, which may offer deeper insights into our understanding of the critical behavior of many-body systems.

\begin{acknowledgments}
The authors would like to thank Ching-Yu Huang for enlightening discussions. This research was supported by Grant No. NSTC 113-2112-M-029-005 and NSTC 113-2112-M-029-006 of the National Science and Technology Council of Taiwan.
\end{acknowledgments}

\appendix
\section{SSE estimators of average work and its temperature derivative}

In this study, we employ the well-established stochastic series expansion (SSE) algorithm~\cite{QMCreview} to compute the reduced average work and its temperature derivative. From Eq.~\eqref{avg_work2}, for a sudden quench in hopping parameter from $t$ to $t+\delta$, the reduced average work $\langle\mathcal{W}\rangle=(1/L^d)\langle W\rangle/\delta$ of a $d$-dimensional system is given by
\begin{equation}\label{ave_W}
\langle\mathcal{W}\rangle = \frac{1}{L^d} \frac{1}{t} \langle H_t \rangle \; .
\end{equation}
Here $\langle O\rangle=\mathrm{Tr}\left(O\,e^{-\beta H}\right)/\mathrm{Tr}\left(e^{-\beta H}\right)$ denotes the thermal average of the operator $O$, and $H_t = -t\sum_{\langle i,j\rangle} b^\dagger_i b_j + \mathrm{h.c.}$ is the kinetic term of the total Hamiltonian $H$. The SSE method expands the exponential operator into a sequence of operators that can be efficiently sampled using diagonal and loop updates. The thermal average of the kinetic energy can be readily obtained in the SSE framework via the relation:
\begin{equation}
\langle H_t \rangle = -\frac{1}{\beta} \langle N_t \rangle.
\end{equation}
Here $N_t$ denotes the number of off-diagonal operators appearing in the operator sequence during sampling~\cite{Sandvik1992,estimators_SSE}.

As discussed in the main text, one can identify the second-order thermal phase transitions by utilizing the temperature derivative of the reduced average work, $\partial\langle\mathcal{W}\rangle/\partial T$, which becomes divergent at critical temperatures in the thermodynamic limit. Direct numerical differentiation to compute $\partial\langle\mathcal{W}\rangle/\partial T$ is problematic, as it can yield inaccurate results due to the statistical nature of $\langle\mathcal{W}\rangle$ obtained from QMC simulations. Fortunately, by differentiating the expression of Eq.~\eqref{ave_W}, this temperature derivative can be formulated as
\begin{equation}
\frac{\partial\langle\mathcal{W}\rangle}{\partial T}
= \frac{1}{L^d} \frac{\beta^2}{t} \left\{ \langle H_t H \rangle - \langle H_t \rangle \langle H \rangle \right\} \; .
\end{equation}
Notice that the thermal averages at the right-handed side can be efficiently measured within the SSE framework. The second term is simply the product of the average kinetic energy and the average total energy, while the first term can be computed using the formula~\cite{Sandvik1992,estimators_SSE}:
\begin{align}
\langle H_t H \rangle = \frac{1}{\beta^2} \langle (n-1) N_t \rangle.
\end{align}
Here $n$ is the total number of operators in the operator sequence. By employing this formula, the temperature derivative of the reduced average work can be calculated precisely while maintaining controlled statistical uncertainties.



\end{document}